\documentclass[11pt]{article}
\usepackage[dvips]{graphicx,psfrag}

\begin{document}
\title{Unification of Neutrino-Neutrino and Neutrino-Antineutrino Oscillations}
\author{
{K. Kimura$^1$}
{and A. Takamura$^{2}$}
\\
\\
\\
{\small \it $^1$Department of Physics, Nagoya University,}
{\small \it Furo-cho, Chikusa-ku, Nagoya, 464-8602, Japan}\\
{\small \it $^2$Department of Mathematics, 
Toyota National Collage of Technology}\\
{\small \it Eisei-cho 2-1, Toyota-shi, 471-8525, Japan}}
\date{}
\maketitle
\vspace{-11cm}
\begin{flushright}
\end{flushright}
\vspace{10.5cm}
\vspace{-2.5cm}
%
%
\vspace{1cm}
\begin{abstract}
In the case of two-generation Majorana neutrinos, we derive the oscillation probabilities for $\nu \leftrightarrow \nu$, $\nu^c \leftrightarrow \nu^c$, and $\nu \leftrightarrow \nu^c$ within a unified framework using a relativistic equation. We demonstrate that the Majorana phase arises not from lepton number violation but from chirality change. Additionally, we verify the conservation of unitarity. Notably, we find that the oscillation probabilities for $\nu \leftrightarrow \nu^c$ derived in this paper differ significantly from previous results, particularly in the absence of zero-distance effects and direct CP violation.
\end{abstract}



\section{Introduction}
\label{sec:intro}

Since the discovery of neutrino oscillations in the Super-Kamiokande atmospheric neutrino experiment in 1998, substantial evidence for these oscillations has accumulated from solar neutrino experiments \cite{SK, SNO, SK2}, long-baseline experiments \cite{T2K, MINOS}, and reactor experiments \cite{KamLAND, DayaBay, RENO, DoubleChooz}. These experiments have precisely determined two mass-squared differences and three mixing angles in the lepton sector. A primary objective of upcoming experiments \cite{HK, DUNE} is to ascertain the leptonic Dirac CP phase and the mass ordering. Accurate estimates of matter effects on neutrino oscillation probabilities have been made to support this goal \cite{Zaglauer, Ohlsson, KTY1, KTY2, Yokomakura0207, Yasuda}.

In the Standard Model, the neutrino is the only electrically neutral fermion and may be a Majorana particle \cite{Majorana}. For Majorana neutrinos, it is known that the Majorana CP phase appears \cite{Majorana-phase, Majorana-phase2, Majorana-phase3, Bernabeu1983}, in addition to the Dirac CP phase. These Majorana CP phases are relevant in frameworks with two or more generations and are thought to manifest in lepton number-violating phenomena such as $0\nu\beta\beta$ decay.

To investigate whether neutrinos are Dirac or Majorana particles, several $0\nu\beta\beta$ decay experiments have been conducted \cite{KamLAND-Zen:2016pfg, Alfonso:2015wka, Albert:2014awa, Agostini:2013mzu, Gando:2012zm, Elliott:2016ble, Andringa:2015tza}, but a definitive conclusion has yet to be reached. In the context of $0\nu\beta\beta$ decay, a neutrino transitions into an antineutrino (the charge conjugate of the neutrino, which has an opposite chirality), and the intermediate state is associated with $\nu_e \leftrightarrow \nu_e^c$ oscillations at atomic scales. The probabilities for these oscillations have been calculated in previous studies \cite{Bahcall1978, Valle1981, Li1982, Bernabeu1983}. In these calculations, only the CP-even effects of the Majorana CP phases are included, leading to equal decay rates for particles and antiparticles, as the flavor of the neutrino does not change. Conversely, it has been noted that the CP-odd effects could be observed through measurements of $\nu_{\alpha} \leftrightarrow \nu_{\beta}^c$ oscillations with different flavors $\alpha$ and $\beta$ \cite{Gouvea2003, Xing2013}.

We have some questions regarding previous studies on $\nu \leftrightarrow \nu^c$ oscillations. When considering $\nu \leftrightarrow \nu^c$ oscillations alongside $\nu \leftrightarrow \nu$ oscillations, the sum of the probabilities deviates from one, as unitarity holds solely within the framework of $\nu \leftrightarrow \nu$ oscillations, as indicated in earlier works. Additionally, reference \cite{Li1982} pointed out that the probability of $\nu \leftrightarrow \nu^c$ oscillations is not zero, even for a zero baseline length, leading to what is known as the zero-distance effect. This finding appears to contradict unitarity as well.

In our previous work \cite{KT1}, we derived the exact oscillation probabilities of Dirac neutrinos relativistically within the two-generation framework. As a result, we achieved a unified understanding of neutrino oscillations, both with and without chirality-flip. We also demonstrated the need for correction terms in the oscillation probabilities to ensure unitarity. If oscillation probabilities can be measured with precision, they may reveal information related to the absolute value of neutrino masses and a new CP phase. This new CP phase influences the oscillation probabilities involving chirality-flip.

In this paper, we aim to comprehensively understand the oscillations between neutrinos and antineutrinos using the relativistic framework established in our previous work. We show that the Majorana CP phase is a specific instance of this new CP phase associated with chirality-flip, similar to the case of Dirac neutrinos discussed in \cite{KT1}. That is, the Majorana CP phase does not arise from lepton number violation but rather from chirality change. In the context of Majorana neutrinos, we interpret the antineutrino $\nu^c$ as serving the role of the right-handed neutrino $\nu_R$ found in Dirac neutrinos, while the new CP phase identified in our prior study becomes the Majorana CP phase.

We also establish that there is no direct CP violation in $\nu_{\alpha} \leftrightarrow \nu_{\beta}^c$ oscillations, even when flavors $\alpha$ and $\beta$ differ. In other words, the difference in CP-conjugate probabilities vanishes, and information about this CP phase can only be inferred indirectly through the cosine term. Furthermore, the zero-distance effect proposed in \cite{Li1982}, which implies a finite oscillation probability from neutrino to antineutrino instantaneously, does not arise from our derived results. These findings represent a significant departure from previous conclusions.

The structure of the paper is organized as follows: In Section II, we define the notation used throughout this work. Section III reviews the relativistic derivation of oscillation probabilities for neutrinos with only the Dirac mass term. In Section IV, we present the derivation of oscillation probabilities for neutrinos with only the Majorana mass term. In Section V, we compare our results with previous findings concerning neutrinos with only the Majorana mass term. Finally, in Section VI, we summarize the results obtained in this paper.

\section{Notation}

In this section, we write down the notation used in this paper. 
We mainly use the chiral representation 
because neutrinos are measured through weak interactions.
In chiral representation, the gamma matrices with $4\times 4$ form are given by 
\begin{eqnarray}
\gamma^0=\left(\begin{array}{cc}0 & 1 \\ 1 & 0\end{array}\right), \qquad 
\gamma^i=\left(\begin{array}{cc}0 & -\sigma_i \\ \sigma_i & 0\end{array}\right), \qquad 
\gamma_5=\left(\begin{array}{cc}1 & 0 \\ 0 & -1\end{array}\right),  \label{gamma-mat}
\end{eqnarray}
where $2\times 2$ $\sigma$ matrices are defined by 
\begin{eqnarray}
\sigma_1=\left(\begin{array}{cc}0 & 1 \\ 1 & 0\end{array}\right), \qquad 
\sigma_2=\left(\begin{array}{cc}0 & -i \\ i & 0\end{array}\right), \qquad 
\sigma_3=\left(\begin{array}{cc}1 & 0 \\ 0 & -1\end{array}\right). 
\end{eqnarray}
We also define 4-component spinors $\psi$, $\psi_L$ and $\psi_R$ as 
\begin{eqnarray}
\psi=\left(\begin{array}{c}\xi \\ \eta \end{array}\right), \qquad 
\psi_L=\frac{1-\gamma_5}{2}\psi=\left(\begin{array}{c}0 \\ \eta\end{array}\right), \qquad 
\psi_R=\frac{1+\gamma_5}{2}\psi=\left(\begin{array}{c}\xi \\ 0 \end{array}\right),   \label{psi-def}
\end{eqnarray}
and 2-component spinors $\xi$ and $\eta$ as 
\begin{eqnarray}
\xi=\left(\begin{array}{c}\nu_R^{\prime} \\ \nu_R \end{array}\right), \qquad 
\eta=\left(\begin{array}{c}\nu_L^{\prime} \\ \nu_L \end{array}\right).  \label{xi-eta}
\end{eqnarray}

We consider the quantum numbers of each component that appears in (\ref{xi-eta}).
We use three operators to distinguish the states of neutrinos: chirality, helicity, and energy.
We calculate the commutation relation between the chirality operator $\gamma_5$, the helicity operator $\Sigma\cdot \hat{p}$, and the Hamiltonian $H$. Here, $\Sigma$ is a matrix arranged in diagonal components with two Pauli matrices, $\hat{p}$ is a unit vector with momentum direction, and
we take the momentum in the $z$ direction.
First, the commutation relation between $\gamma_5$ and $H$ is
\begin{eqnarray}
[\gamma_5, H]&=&[\gamma_5, {\bf \alpha}\cdot p+\beta m]=[\gamma_5, \gamma^0\gamma^i p_i+\gamma^0 m]. 
\nonumber \\\
&=&\gamma_5(\gamma^0\gamma^i p_i+\gamma^0 m)-(\gamma^0\gamma^i p_i+\gamma^0 m)\gamma_5
=2\gamma_5\gamma^0 m\neq 0, 
\end{eqnarray} 
and does not commute. 
This means that the left-handed neutrino produced as an eigenstate of chirality can change over time and become a right-handed neutrino.

Second, the commutation relation between $\gamma_5$ and the helicity operator $\Sigma\cdot \hat{p}$ is
\begin{eqnarray}
[\gamma_5, \Sigma\cdot \hat{p}]&=&\left(\begin{array}{cc}
1 & 0 \\
0 & -1 
\end{array}\right)
\left(\begin{array}{cc}
\sigma\cdot \hat{p} & 0 \\ 
0 & \sigma\cdot \hat{p} 
\end{array}\right)-\left(\begin{array}{cc}
\sigma\cdot \hat{p} & 0 \\ 
0 & \sigma\cdot \hat{p} 
\end{array}\right)\left(\begin{array}{cc}
1 & 0 \\
0 & -1 
\end{array}\right)
=0
\end{eqnarray} 
This means that chirality and helicity can be diagonalized simultaneously. 

Finally, the commutation relation between helicity and Hamiltonian is well known as
\begin{eqnarray}
[H, \Sigma\cdot \hat{p}]&=&
\left[
\left(\begin{array}{cc}
\sigma\cdot p & 0 \\ 
0 & -\sigma\cdot p 
\end{array}\right)+\left(\begin{array}{cc}
0 & m \\
m & 0 
\end{array}\right),
\left(\begin{array}{cc}
\sigma\cdot \hat{p} & 0 \\ 
0 & \sigma\cdot \hat{p} 
\end{array}\right)\right]
=0
\end{eqnarray}
This means that the helicity is conserved and can be diagonalized simultaneously with the Hamiltonian. 
This is also consistent with the conservation law of angular momentum.

Taking into account that neutrinos are produced through weak interactions, we take the initial state of the neutrino to be an eigenstate of chirality and helicity.
Since neutrinos have a very small but finite mass, there are states with both positive and negative helicity 
for each neutrino with left-handed and right-handed chirality.
For neutrinos produced with certain chirality and helicity, the chirality changes with time, but its helicity does not.
Also note that in this case, the initial state of the neutrino is not an energy eigenstate.

As we can see by multiplying the chirality and helicity operators
to the neutrino state, the components, $\nu_L$ and $\nu_R$ in (\ref{xi-eta}), are states with negative helicity, as shown in the Table 1 , and $\nu_L^{\prime}$ and $\nu_R^{\prime}$ have positive helicity.
Also, antineutrinos are defined as charge conjugation of neutrinos, so their momentum is opposite to that of neutrinos.
For this reason, the helicity of antineutrinos is of the opposite sign to that of neutrinos, even though their spin direction is the same.

\begin{table}[tb]
  \begin{center} 
    \caption{chirality and helicity of each component}
\begin{tabular}{|c|c|c|c|c|c|c|c|c|}
\hline
& $\nu_R^{\prime}$ & $\nu_R$ & $\nu_L^{\prime}$ & $\nu_L$ 
& $\nu_R^{c\prime}$ & $\nu_R^c$ & $\nu_L^{c\prime}$ & $\nu_L^c$\\\hline
chirality & Right & Right & Left & Left & Left & Left & Right & Right \\\hline
helicity & $+$ & $-$ & $+$ & $-$ & $-$ & $+$ & $-$ & $+$
\\\hline
\end{tabular}
  \end{center}
\end{table}

Furthermore, 
we use the subscript $\alpha$ and $\beta$ for flavor, $L$ and $R$ for chirality, 
the number $1$ and $2$ for generation and superscript $\pm$ for energy. 
Because of negligible neutrino mass, mass eigenstate has been often identified with energy eigenstate 
in many papers.
But in the future, we should distinguish these two kinds of eigenstates because of the finite neutrino mass. 
More concretely, we use the following eigenstates; 
\begin{eqnarray}
&&{\rm flavor \,\, eigenstates}: \quad \nu_{\alpha L}, \nu_{\alpha R}, \nu_{\beta L}, \nu_{\beta R}, \\
&&{\rm mass \,\, eigenstates}: \quad\,\, \nu_{1L}, \nu_{1R}, \nu_{2L}, \nu_{2R}, \\
&&{\rm energy \,\, eigenstates}: \quad  \,\,\,\nu_1^+, \nu_1^-, \nu_2^+, \nu_2^-.
\end{eqnarray}
It is noted that mass eigenstates are not exactly the eigenstates.
We use the term, eigenstates, in the sense that the mass submatrix in the Hamiltonian is diagonalized.
We also difine the spinors for antineutrino as charge conjugation of neutrino $\psi^c=i\gamma^2 \psi^*$.
The charge conjugations for left-handed and right-handed neutrinos are defined by 
\begin{eqnarray}
\psi_L^c\equiv (\psi_L)^c\!\!\equiv\!\!
\left(\begin{array}{c}\nu_L^c \\ \nu_L^{c\prime} \\ 0 \\ 0 \end{array}\right) 
\!\!\equiv i\gamma^2 \psi_L^*\!=\!i\gamma^2 \frac{1-\gamma_5}{2}\psi^*\! 
=\!\frac{1+\gamma_5}{2}(i\gamma^2 \psi^*)
\!=\!\!(\psi^c)_R\!\!=\!\left(\!\!\begin{array}{c}i\sigma_2 \eta^* \\ 0 \end{array}\!\!\right)
\!=\!\left(\!\!\begin{array}{c}\nu_L^* \\ -\nu_L^{*\prime} \\ 0 \\ 0 \end{array}\!\!\right), \label{nuc} 
\end{eqnarray}
\begin{eqnarray}
\psi_R^c\equiv (\psi_R)^c\!\!\equiv\!\!\left(\begin{array}{c}0 \\ 0 \\ \nu_R^c \\ \nu_R^{c\prime} \end{array}\right)
\!\!\equiv i\gamma^2 \psi_R^*\!=\!i\gamma^2 \frac{1+\gamma_5}{2}\psi^*\! 
=\!\frac{1-\gamma_5}{2}(i\gamma^2 \psi^*)
\!=\!\!(\psi^c)_L\!\!=\!\left(\!\!\begin{array}{c}0 \\ -i\sigma_2 \xi^* \end{array}\!\!\right)
\!=\!\left(\!\!\begin{array}{c}0 \\ 0 \\ -\nu_R^* \\ \nu_R^{*\prime} \end{array}\!\!\right). 
\end{eqnarray}
It is noted that the chirality is flipped by taking the charge conjugation.

\section{Review of Dirac Neutrino Oscillation Probabilities in two generations}

In our previous paper \cite{KT1}, we derived the neutrino oscillation probabilities in the case of 
the Dirac neutrinos by using a relativistic equation, the Dirac equation.
In this section, we review the results of our previous paper \cite{KT1} in order to clarify the similarity and 
the difference between the case of the Dirac neutrinos and the Majorana neutrinos. 

In the case of two generations, the Lagrangian of neutrinos with the Dirac mass term is given by 
\begin{eqnarray}
L&=&i\overline{\psi_{\alpha L}}\gamma^\mu \partial_\mu \psi_{\alpha L}+i\overline{\psi_{\alpha R}}\gamma^\mu \partial_\mu \psi_{\alpha R}
+i\overline{\psi_{\beta L}}\gamma^\mu \partial_\mu \psi_{\beta L}+i\overline{\psi_{\beta R}}\gamma^\mu \partial_\mu \psi_{\beta R}\nonumber \\
&&-\left[\overline{\psi_{\alpha L}}m_{\alpha\alpha}^*\psi_{\alpha R}+\overline{\psi_{\beta L}}m_{\beta\beta}^*\psi_{\beta R}
+\overline{\psi_{\alpha L}}m_{\beta\alpha}^*\psi_{\beta R}
+\overline{\psi_{\beta L}}m_{\alpha\beta}^*\psi_{\alpha R}\right] \nonumber \\
&&-\left[\overline{\psi_{\alpha R}}m_{\alpha\alpha}\psi_{\alpha L}+\overline{\psi_{\beta R}}m_{\beta\beta}\psi_{\beta L}
+\overline{\psi_{\alpha R}}m_{\alpha\beta}\psi_{\beta L}
+\overline{\psi_{\beta R}}m_{\beta\alpha}\psi_{\alpha L}\right],
\end{eqnarray}
where $m_{\alpha\beta}$ etc. stands for the Dirac masses.
The Eular-Lagrange equations for $\overline{\psi_{\alpha L}}$, $\overline{\psi_{\alpha R}}$, 
$\overline{\psi_{\beta L}}$ and $\overline{\psi_{\beta R}}$ are combined to the matrix form, 
\begin{eqnarray}
i\frac{d}{dt}\left(\begin{array}{c}
\nu_{\alpha R}^{\prime} \\ \nu_{\beta R}^{\prime} \\ \nu_{\alpha L}^{\prime} \\ \nu_{\beta L}^{\prime} \\
\nu_{\alpha R} \\ \nu_{\beta R} \\ \nu_{\alpha L} \\ \nu_{\beta L}
\end{array}\right)
=\left(\begin{array}{cccc|cccc}
p & 0 & m_{\alpha\alpha} & m_{\alpha\beta} & 0 & 0 & 0 & 0 \\ 
0 & p & m_{\beta\alpha} & m_{\beta\beta} & 0 & 0 & 0 & 0 \\
m_{\alpha\alpha}^* & m_{\beta\alpha}^* & -p & 0 & 0 & 0 & 0 & 0 \\ 
m_{\alpha\beta}^* & m_{\beta\beta}^* & 0 & -p & 0 & 0 & 0 & 0 \\
\hline
0 & 0 & 0 & 0 & -p & 0 & m_{\alpha\alpha} & m_{\alpha\beta} \\
0 & 0 & 0 & 0 & 0 & -p & m_{\beta\alpha} & m_{\beta\beta} \\
0 & 0 & 0 & 0 & m_{\alpha\alpha}^* & m_{\beta\alpha}^* & p & 0 \\
0 & 0 & 0 & 0 & m_{\alpha\beta}^* & m_{\beta\beta}^* & 0 & p 
\end{array}\right)\left(\begin{array}{c}
\nu_{\alpha R}^{\prime} \\ \nu_{\beta R}^{\prime} \\ \nu_{\alpha L}^{\prime} \\ \nu_{\beta L}^{\prime} \\
\nu_{\alpha R} \\ \nu_{\beta R} \\ \nu_{\alpha L} \\ \nu_{\beta L}
\end{array}\right),
\end{eqnarray}
where we take equal momentum assumption for neutrinos with different flavors, see ref. \cite{KT1} for details. 
In the above equation, the bottom-right part and the top-left part are completely 
separated and therefore they do not mix each other even if the time has passed.
Taking the bottom right part, we obtain the equation,  
\begin{eqnarray}
i\frac{d}{dt}\left(\begin{array}{c}
\nu_{\alpha R} \\ \nu_{\beta R} \\ \nu_{\alpha L} \\ \nu_{\beta L}
\end{array}\right)
=\left(\begin{array}{cc|cc}
-p & 0 & m_{\alpha\alpha} & m_{\alpha\beta} \\
0 & -p & m_{\beta\alpha} & m_{\beta\beta} \\\hline
m_{\alpha\alpha}^* & m_{\beta\alpha}^* & p & 0 \\
m_{\alpha\beta}^* & m_{\beta\beta}^* & 0 & p 
\end{array}\right)\left(\begin{array}{c}
\nu_{\alpha R} \\ \nu_{\beta R} \\ \nu_{\alpha L} \\ \nu_{\beta L}
\end{array}\right), \label{Dirac-Hamiltonian}
\end{eqnarray}
where $m_{\alpha\alpha}$, $m_{\beta\beta}$ and $m_{\alpha\beta}$ are complex in general.
The flavor eigenstates are represented as the linear combination of the mass eigenstates,  
\begin{eqnarray}
\left(\begin{array}{c}
\nu_{\alpha R} \\ \nu_{\beta R} \\ \nu_{\alpha L} \\ \nu_{\beta L}
\end{array}\right)
=\left(\begin{array}{cc|cc}
V_{\alpha 1} & V_{\alpha 2} & 0 & 0 \\
V_{\beta 1} & V_{\beta 2} & 0 & 0 \\
\hline
0 & 0 & U_{\alpha 1} & U_{\alpha 2} \\
0 & 0 & U_{\beta 1} & U_{\beta 2} 
\end{array}\right)\left(\begin{array}{c}
\nu_{1R} \\ \nu_{2R} \\ \nu_{1L} \\ \nu_{2L}
\end{array}\right). \label{CFstate to CMstate}
\end{eqnarray}
The Dirac mass term of the Hamiltonian in eq.(\ref{Dirac-Hamiltonian}) is diagonalized by the above mixing matrix as shown in 
\begin{eqnarray}
\hspace{-0.7cm}
\left(\begin{array}{cc|cc}
V_{\alpha 1}^* & V_{\beta 1}^* & 0 & 0 \\
V_{\alpha 2}^* & V_{\beta 2}^* & 0 & 0 \\
\hline
0 & 0 & U_{\alpha 1}^* & U_{\beta 1}^* \\
0 & 0 & U_{\alpha 2}^* & U_{\beta 2}^* 
\end{array}\right)\!\!\!
\left(\!\!\begin{array}{cccc}
-p & 0 & m_{\alpha\alpha} & m_{\alpha\beta} \\
0 & -p & m_{\beta\alpha} & m_{\beta\beta} \\
m_{\alpha\alpha}^* & m_{\beta\alpha}^* & p & 0 \\
m_{\alpha\beta}^* & m_{\beta\beta}^* & 0 & p 
\end{array}\!\!\right)\!\!\!
\left(\begin{array}{cc|cc}
V_{\alpha 1} & V_{\alpha 2} & 0 & 0 \\
V_{\beta 1} & V_{\beta 2} & 0 & 0 \\
\hline
0 & 0 & U_{\alpha 1} & U_{\alpha 2} \\
0 & 0 & U_{\beta 1} & U_{\beta 2} 
\end{array}\right)\!\!
=\!\!\left(\!\!\begin{array}{cccc}
-p & 0 & m_1 & 0 \\
0 & -p & 0 & m_2 \\
m_1 & 0 & p & 0 \\
0 & m_2 & 0 & p 
\end{array}\!\!\right)\!\!,
\end{eqnarray}
and the time evolution of the mass eigenstates is given by 
\begin{eqnarray}
i\frac{d}{dt}\left(\begin{array}{c}
\nu_{1R} \\ \nu_{2R} \\ \nu_{1L} \\ \nu_{2L}
\end{array}\right)
=\left(\begin{array}{cccc}
-p & 0 & m_1 & 0 \\
0 & -p & 0 & m_2 \\
m_1 & 0 & p & 0 \\
0 & m_2 & 0 & p 
\end{array}\right)\left(\begin{array}{c}
\nu_{1R} \\ \nu_{2R} \\ \nu_{1L} \\ \nu_{2L}
\end{array}\right), \label{CMstate eq}
\end{eqnarray}
where $m_1$ and $m_2$ are mass eigenvalues of neutrinos.
Furthermore, we rewrite the above equations to those for the energy eigenstates.
Exchanging some rows and some columns in eq.(\ref{CMstate eq}), we obtain 
\begin{eqnarray}
i\frac{d}{dt}\left(\begin{array}{c}
\nu_{1R} \\ \nu_{1L} \\ \nu_{2R} \\ \nu_{2L}
\end{array}\right)
=\left(\begin{array}{cc|cc}
-p & m_1 & 0 & 0 \\
m_1 & p & 0 & 0 \\
\hline  
0 & 0 & -p & m_2 \\
0 & 0 & m_2 & p 
\end{array}\right)\left(\begin{array}{c}
\nu_{1R} \\ \nu_{1L} \\ \nu_{2R} \\ \nu_{2L}
\end{array}\right).
\end{eqnarray}
The mass eigenstates are represented by 
the linear combination of the energy eigenstates, 
\begin{eqnarray}
\left(\begin{array}{c}
\nu_{1R} \\ \nu_{1L} \\ \nu_{2R} \\ \nu_{2L}
\end{array}\right)
=\left(\begin{array}{cc|cc}
\sqrt{\frac{E_1+p}{2E_1}} & \sqrt{\frac{E_1-p}{2E_1}} & 0 & 0 \\
-\sqrt{\frac{E_1-p}{2E_1}} & \sqrt{\frac{E_1+p}{2E_1}} & 0 & 0 \\
\hline
0 & 0 & \sqrt{\frac{E_2+p}{2E_2}} & \sqrt{\frac{E_2-p}{2E_2}} \\
0 & 0 & -\sqrt{\frac{E_2-p}{2E_2}} & \sqrt{\frac{E_2+p}{2E_2}} 
\end{array}\right)\left(\begin{array}{c}
\nu_{1}^- \\ \nu_{1}^+ \\ \nu_{2}^- \\ \nu_{2}^+
\end{array}\right), \label{CMstate to EHstate}
\end{eqnarray}
and the Hamiltonian in eq.(\ref{CMstate eq}) is completely diagonalized by the mixing matrix in (\ref{CMstate to EHstate}) and 
the time evolution of the energy eigenstates is given by 
\begin{eqnarray}
i\frac{d}{dt}\left(\begin{array}{c}
\nu_{1}^- \\ \nu_{1}^+ \\ \nu_{2}^- \\ \nu_{2}^+
\end{array}\right)
=\left(\begin{array}{cccc}
-E_1 & 0 & 0 & 0 \\
0 & E_1 & 0 & 0 \\
0 & 0 & -E_2 & 0 \\
0 & 0 & 0 & E_2 
\end{array}\right)\left(\begin{array}{c}
\nu_{1}^- \\ \nu_{1}^+ \\ \nu_{2}^- \\ \nu_{2}^+
\end{array}\right), 
\end{eqnarray}
where $E_1$ and $E_2$ are the neutrino energies given by 
\begin{eqnarray}
E_1=\sqrt{p^2+m_1^2}, \qquad 
E_2=\sqrt{p^2+m_2^2}. 
\end{eqnarray}
Combining eq. (\ref{CFstate to CMstate}) and (\ref{CMstate to EHstate}), 
the flavor eigenstates are represented by the energy eigenstates as
\begin{eqnarray}
\left(\begin{array}{c}
\nu_{\alpha R} \\ \nu_{\beta R} \\ \nu_{\alpha L} \\ \nu_{\beta L}
\end{array}\right)\!\!\!
&=&\!\!\!
\left(\begin{array}{cc|cc}
V_{\alpha 1} & V_{\alpha 2} & 0 & 0 \\
V_{\beta 1} & V_{\beta 2} & 0 & 0 \\
\hline
0 & 0 & U_{\alpha 1} & U_{\alpha 2} \\
0 & 0 & U_{\beta 1} & U_{\beta 2} 
\end{array}\right)\!\!\!
\left(\begin{array}{cccc}
1 & 0 & 0 & 0 \\
0 & 0 & 1 & 0 \\
0 & 1 & 0 & 0 \\
0 & 0 & 0 & 1 
\end{array}\right)\!\!\!
\left(\begin{array}{cc|cc}
\sqrt{\frac{E_1+p}{2E_1}} & \sqrt{\frac{E_1-p}{2E_1}} & 0 & 0 \\
-\sqrt{\frac{E_1-p}{2E_1}} & \sqrt{\frac{E_1+p}{2E_1}} & 0 & 0 \\
\hline
0 & 0 & \sqrt{\frac{E_2+p}{2E_2}} & \sqrt{\frac{E_2-p}{2E_2}} \\
0 & 0 & -\sqrt{\frac{E_2-p}{2E_2}} & \sqrt{\frac{E_2+p}{2E_2}} 
\end{array}\right)\!\!\!\left(\begin{array}{c}
\nu_{1}^- \\ \nu_{1}^+ \\ \nu_{2}^- \\ \nu_{2}^+
\end{array}\right) \nonumber \\
&=&\left(\begin{array}{cc|cc}
\sqrt{\frac{E_1+p}{2E_1}}V_{\alpha 1} & \sqrt{\frac{E_1-p}{2E_1}}V_{\alpha 1} 
& \sqrt{\frac{E_2+p}{2E_2}}V_{\alpha 2} & \sqrt{\frac{E_2-p}{2E_2}}V_{\alpha 2} \\
\sqrt{\frac{E_1+p}{2E_1}}V_{\beta 1} & \sqrt{\frac{E_1-p}{2E_1}}V_{\beta 1} 
& \sqrt{\frac{E_2+p}{2E_2}}V_{\beta 2} & \sqrt{\frac{E_2-p}{2E_2}}V_{\beta 2} \\
\hline
-\sqrt{\frac{E_1-p}{2E_1}}U_{\alpha 1} & \sqrt{\frac{E_1+p}{2E_1}}U_{\alpha 1} 
& -\sqrt{\frac{E_2-p}{2E_2}}U_{\alpha 2} & \sqrt{\frac{E_2+p}{2E_2}}U_{\alpha 2} \\
-\sqrt{\frac{E_1-p}{2E_1}}U_{\beta 1} & \sqrt{\frac{E_1+p}{2E_1}}U_{\beta 1} 
& -\sqrt{\frac{E_2-p}{2E_2}}U_{\beta 2} & \sqrt{\frac{E_2+p}{2E_2}}U_{\beta 2} 
\end{array}\right)
\left(\begin{array}{c}
\nu_{1}^- \\ \nu_{1}^+ \\ \nu_{2}^- \\ \nu_{2}^+
\end{array}\right). 
\end{eqnarray}
After rewriting the equations for fields to those for one particle states, 
we can calculate the oscillation amplitudes for $\nu_{\alpha L}$ to other neutrinos,
\begin{eqnarray}
&&\hspace{-1cm}A(\nu_{\alpha L}\to\nu_{\alpha L})=\langle \nu_{\alpha L}|\nu_{\alpha L}(t)\rangle \nonumber \\
&&\hspace{-0.8cm}=|U_{\alpha 1}|^2\cos (E_1t)-i|U_{\alpha 1}|^2\cdot \frac{p}{E_1}\sin (E_1t)
+|U_{\alpha 2}|^2\cos (E_2t)-i|U_{\alpha 2}|^2\cdot \frac{p}{E_2}\sin (E_2t), \\
&&\hspace{-1cm}A(\nu_{\alpha L}\to\nu_{\beta L})=\langle \nu_{\beta L}|\nu_{\alpha L}(t)\rangle \nonumber \\
&&\hspace{-0.8cm}=U_{\alpha 1}^*U_{\beta 1}\left\{\cos (E_1t)-i\frac{p}{E_1}\sin (E_1t)\right\}
+U_{\alpha 2}^*U_{\beta 2}\left\{\cos (E_2t)-i\frac{p}{E_2}\sin (E_2t)\right\}, \\
&&\hspace{-1cm}A(\nu_{\alpha L}\to\nu_{\alpha R})=\langle \nu_{\alpha R}|\nu_{\alpha L}(t)\rangle
=-iU_{\alpha 1}^*V_{\alpha 1}\frac{m_1}{E_1}\sin (E_1t)
-iU_{\alpha 2}^*V_{\alpha 2}\frac{m_2}{E_2}\sin (E_2t), \\
&&\hspace{-1cm}A(\nu_{\alpha L}\to\nu_{\beta R})=\langle \nu_{\beta R}|\nu_{\alpha L}(t)\rangle
=-iU_{\alpha 1}^*V_{\beta 1}\frac{m_1}{E_1}\sin (E_1t)
-iU_{\alpha 2}^*V_{\beta 2}\frac{m_2}{E_2}\sin (E_2t). 
\end{eqnarray}
Furthermore, we also obtain the oscillation probabilities for $\nu_{\alpha L}$ by squaring the amplitudes as 
\begin{eqnarray}
&&\hspace{-0.8cm}P(\nu_{\alpha L}\to\nu_{\alpha L})
=\{|U_{\alpha 1}|^2\cos (E_1t)+|U_{\alpha 2}|^2\cos (E_2t)\}^2 \nonumber \\
&&\hspace{-0.3cm}+\left\{|U_{\alpha 1}|^2\cdot \frac{p}{E_1}\sin (E_1t)
+|U_{\alpha 2}|^2\cdot \frac{p}{E_2}\sin (E_2t)\right\}^2, \label{P(aL-to-aL)}\\
&&\hspace{-0.8cm}P(\nu_{\alpha L}\to\nu_{\beta L})=
|U_{\alpha 1}U_{\beta 1}|^2\left\{\cos^2 (E_1t)+\frac{p^2}{E_1^2}\sin^2 (E_1t)\right\} \nonumber \\
&&\hspace{-0.3cm}+|U_{\alpha 2}U_{\beta 2}|^2\left\{\cos^2 (E_2t)+\frac{p^2}{E_2^2}\sin^2 (E_2t)\right\} \nonumber \\
&&\hspace{-0.3cm}+2{\rm Re}\left[U_{\alpha 1}^*U_{\beta 1}\left\{\cos (E_1t)-i\frac{p}{E_1}\sin (E_1t)\right\}
U_{\alpha 2}U_{\beta 2}^*\left\{\cos (E_2t)+i\frac{p}{E_2}\sin (E_2t)\right\}\right], 
\end{eqnarray}
\begin{eqnarray}
&&\hspace{-0.8cm}P(\nu_{\alpha L}\to\nu_{\alpha R})=
|U_{\alpha 1}V_{\alpha 1}|^2\frac{m_1^2}{E_1^2}\sin^2 (E_1t)+|U_{\alpha 2}V_{\alpha 2}|^2\frac{m_2^2}{E_2^2}\sin^2 (E_2t) \nonumber \\
&&\hspace{-0.3cm}+2{\rm Re}[U_{\alpha 1}U_{\alpha 2}^*V_{\alpha 1}^*V_{\alpha 2}]\frac{m_1m_2}{E_1E_2}\sin (E_1t)\sin (E_2t), \\
&&\hspace{-0.8cm}P(\nu_{\alpha L}\to\nu_{\beta R})=
|U_{\alpha 1}V_{\beta 1}|^2\frac{m_1^2}{E_1^2}\sin^2 (E_1t)
+|U_{\alpha 2}V_{\beta 2}|^2\frac{m_2^2}{E_2^2}\sin^2 (E_2t) \nonumber \\
&&\hspace{-0.3cm}+2{\rm Re}[U_{\alpha 1}U_{\alpha 2}^*V_{\beta 1}^*V_{\beta 2}]\frac{m_1m_2}{E_1E_2}\sin (E_1t)\sin (E_2t). \label{P(aL-to-bR)}
\end{eqnarray}
The $2\times 2$ unitary matrix has four parameters in general, and $U$ and $V$ can be parametrized as 
\begin{eqnarray}
\hspace{-0.5cm}U\!\!&=&\!\!\left(\!\!\begin{array}{cc}
e^{i\rho_{1L}} & 0 \\ 
0 & e^{i\rho_{2L}} 
\end{array}\!\!\right)\!\!
\left(\begin{array}{cc}
\cos \theta_L & \sin \theta_L \\ 
-\sin \theta_L & \cos \theta_L 
\end{array}\right)\!\!\left(\begin{array}{cc}
1 & 0 \\ 
0 & e^{i\phi_L} 
\end{array}\!\!\right)
\!\!=\!\!
\left(\begin{array}{cc}
e^{i\rho_{1L}}\cos \theta_L & e^{i(\rho_{1L}+\phi_L)}\sin \theta_L \\ 
-e^{i\rho_{2L}}\sin \theta_L & e^{i(\rho_{2L}+\phi_L)}\cos \theta_L 
\end{array}\!\!\right),
\\
\hspace{-0.5cm}V\!\!&=&\!\!\left(\!\!\begin{array}{cc}
e^{i\rho_{1R}} & 0 \\ 
0 & e^{i\rho_{2R}} 
\end{array}\!\!\right)\!\!
\left(\begin{array}{cc}
\cos \theta_R & \sin \theta_R \\ 
-\sin \theta_R & \cos \theta_R 
\end{array}\right)\!\!\left(\begin{array}{cc}
1 & 0 \\ 
0 & e^{i\phi_R} 
\end{array}\!\!\right)\!\!
=\!\!
\left(\!\!\begin{array}{cc}
e^{i\rho_{1R}}\cos \theta_R & e^{i(\rho_{1R}+\phi_R)}\sin \theta_R \\ 
-e^{i\rho_{2R}}\sin \theta_R & e^{i(\rho_{2R}+\phi_R)}\cos \theta_R 
\end{array}\!\!\right),
\end{eqnarray}
respectively.
Substituting these parametrizations into the equations (\ref{P(aL-to-aL)})-(\ref{P(aL-to-bR)}), the oscillation probabilities 
can be rewritten as 
\begin{eqnarray}
&&\hspace{-0.5cm}P(\nu_{\alpha L}\to\nu_{\alpha L})=
1-4s_L^2c_L^2\sin^2 \frac{(E_2-E_1)t}{2} \label{old-survive} \\
&&\hspace{-0.5cm}-\left[c_L^4\cdot \frac{m_1^2}{E_1^2}\sin^2 (E_1t)
+s_L^4\cdot \frac{m_2^2}{E_2^2}\sin^2 (E_2t)+2s_L^2c_L^2 \left(\!1-\frac{p^2}{E_1E_2}\right)
\sin (E_1t)\sin (E_2t)\right]\!, \label{new-survive} \\
&&\hspace{-0.5cm}P(\nu_{\alpha L}\to\nu_{\beta L})=
4s_L^2c_L^2\sin^2 \frac{(E_2-E_1)t}{2} \label{old-transition} \\
&&\hspace{0cm}-s_L^2c_L^2\left[\frac{m_1^2}{E_1^2}\sin^2 (E_1t)+\frac{m_2^2}{E_2^2}\sin^2 (E_2t)
-2\left(1-\frac{p^2}{E_1E_2}\right)\sin (E_1t)\sin (E_2t)\right], \label{new-transition} \\
&&\hspace{-0.5cm}P(\nu_{\alpha L}\to\nu_{\alpha R})=
c_L^2c_R^2\frac{m_1^2}{E_1^2}\sin^2 (E_1t)+s_L^2s_R^2\frac{m_2^2}{E_2^2}\sin^2 (E_2t) \nonumber \\
&&\hspace{2cm}+2s_Ls_Rc_Lc_R\cos (\phi_R-\phi_L)\frac{m_1m_2}{E_1E_2}\sin (E_1t)\sin (E_2t), \label{h-change-survive} \\
&&\hspace{-0.5cm}P(\nu_{\alpha L}\to\nu_{\beta R})=
c_L^2s_R^2\frac{m_1^2}{E_1^2}\sin^2 (E_1t)
+s_L^2c_R^2\frac{m_2^2}{E_2^2}\sin^2 (E_2t) \nonumber \\
&&\hspace{2cm}-2s_Ls_Rc_Lc_R\cos (\phi_R-\phi_L)\frac{m_1m_2}{E_1E_2}\sin (E_1t)\sin (E_2t), \label{h-change-transition}
\end{eqnarray}
where we use the abbreviation $\sin \theta_L=s_L$, $\cos \theta_L=c_L$ and so on.
Although the oscillation probabilities for both cases, with and without chirality-flip have been derived separately 
in the calculations done so far, we can derive both types of oscillation probabilities in a unified framework 
by using the Dirac equation.

Let us describe some important points in these oscillation probabilities.
First, we find the new terms (\ref{new-survive}) and (\ref{new-transition}) with an order of $O(m^2/E^2)$ 
in the probabilities without chirality-flip in addition to the well-known terms 
(\ref{old-survive}) and (\ref{old-transition}).
On the other hand, the probabilities with chirality-flip (\ref{h-change-survive}) and (\ref{h-change-transition}) 
are also an order of $O(m^2/E^2)$ 
and the new terms with  $O(m^2/E^2)$ cancel each other out. 
Therefore, the sum of the four probabilities is kept at one.
Second, these new terms depend on not only mass squared differences but also the absolute masses of neutrinos.
This fact is not clearly insisted on previous papers as far as we know.
Third, if we can distinguish the flavor of right-handed particles in physics beyond the Standard Model, 
a new CP phase appears even in two-generation Dirac neutrinos.
This new phase is included in the probabilities with chirality-flip.

Although it is considered that the CP phase does not appear in two-generation Dirac neutrino oscillations, 
we have shown the dependence of the probabilities on the new mixing angle and the new CP phase by considering the left-handed and right-handed neutrinos in the same framework.
These effects are proportional to $(m/E)^2$ and are tiny in usual neutrino oscillation experiments.
If $\nu_R$ do not have weak interactions as in the Standard Model, 
and we can choose the weak eigenstates to be the same as the mass eigenstates, 
we cannot observe the CP phase accompanying the unitary matrix $V$. 
However, in physics beyond the Standard Model, 
if we can distinguish the flavor of right-handed neutrinos, the effects may be measurable in the oscillations of atomic size \cite{KT1}.

If neutrino is the Majorana particle, $\nu_L^{c}$ takes the place of $\nu_R$. 
In this case, also $\nu_L^c$ has weak interactions and we can distinguish the weak eigenstates from mass eigenstates.
Thus, this is one of the examples for the phase $\phi_R-\phi_L$ to be observable. 
In the next section, we derive the oscillation probabilities of the Majorana neutrinos and 
we show that the phase $\phi_R-\phi_L$ actually becomes observable.

\section{Majorana Neutrino Oscillation Probabilities in two generations}

In this section, we consider the transition of two-generation Majorana neutrinos in vacuum.
Let us start from the lagrangian of the Majorana neutrinos, 
\begin{eqnarray}
L&=&\frac{1}{2}i\overline{\psi_{\alpha L}}\gamma^\mu \partial_\mu \psi_{\alpha L}
+\frac{1}{2}i\overline{\psi_{\alpha L}^c}\gamma^\mu \partial_\mu \psi_{\alpha L}^c
+\frac{1}{2}i\overline{\psi_{\beta L}}\gamma^\mu \partial_\mu \psi_{\beta L}
+\frac{1}{2}i\overline{\psi_{\beta L}^c}\gamma^\mu \partial_\mu \psi_{\beta L}^c \nonumber \\
&&-\frac{1}{2}\left[\overline{\psi_{\alpha L}^c}M_{\alpha\alpha}\psi_{\alpha L}
+\overline{\psi_{\beta L}^c}M_{\beta\beta}\psi_{\beta L}
+\overline{\psi_{\alpha L}^c}M_{\alpha\beta}\psi_{\beta L}
+\overline{\psi_{\beta L}^c}M_{\beta\alpha}\psi_{\alpha L}\right]\nonumber \\
&&-\frac{1}{2}\left[\overline{\psi_{\alpha L}}M_{\alpha\alpha}^*\psi_{\alpha L}^c
+\overline{\psi_{\beta L}}M_{\beta\beta}^*\psi_{\beta L}^c
+\overline{\psi_{\beta L}}M_{\alpha\beta}^*\psi_{\alpha L}^c
+\overline{\psi_{\alpha L}}M_{\beta\alpha}^*\psi_{\beta L}^c
\right].
\end{eqnarray}
About the kinetic terms, we have the relation,   
\begin{eqnarray}
{\cal L}_{\rm kin}=i\overline{\psi_{\alpha L}}\gamma^\mu \partial_\mu \psi_{\alpha L}
=i\overline{\psi_{\alpha L}^c}\gamma^\mu \partial_\mu \psi_{\alpha L}^c.
\end{eqnarray}
So, the Eular-Lagrange equation for $\overline{\psi_{\alpha L}}$,  
\begin{eqnarray}
\frac{\partial L}{\partial \overline{\psi_{\alpha L}}}
-\partial_\mu \left(\frac{\partial L}{\partial(\partial_\mu \overline{\psi_{\alpha L}})}\right)=0
\end{eqnarray}
provides 
\begin{eqnarray}
i\gamma^\mu \partial_\mu \psi_{\alpha L}-M_{\alpha\alpha}^* \psi_{\alpha L}^c-M_{\beta\alpha}^* \psi_{\beta L}^c=0,
\end{eqnarray}
where the factor $\frac{1}{2}$ of the kinetic terms and the mass terms dissapears 
because $\overline{\psi_{\alpha L}}$ is regarded as same as $\psi_{\alpha}^c$.
Multiplying $\gamma_0$ from the left, we obtain the equation,  
\begin{eqnarray}
&&i\partial_0 \psi_{\alpha L}+i\gamma^0\gamma^i\partial_i \psi_{\alpha L}
-M_{\alpha\alpha}^* \gamma^0\psi_{\alpha L}^c-M_{\beta\alpha}^* \gamma^0\psi_{\beta L}^c=0.
\end{eqnarray}
Substituting (\ref{psi-def}) and (\ref{nuc}) into this equation, 
we obtain the equation for two-component spinor $\eta_{\alpha}$, 
\begin{eqnarray}
i\partial_0 \left(\begin{array}{c}0 \\ \eta_{\alpha}\end{array}\right)-i
\left(\begin{array}{c}0 \\ \sigma_i\partial_i\eta_{\alpha}\end{array}\right)
-M_{\alpha\alpha}^* \left(\begin{array}{c}0 \\ i\sigma_2\eta_{\alpha}^*\end{array}\right)
-M_{\beta\alpha}^* \left(\begin{array}{c}0 \\ i\sigma_2\eta_{\beta}^*\end{array}\right)=0. 
\end{eqnarray}
Taking out the lower two components, we obtain 
\begin{eqnarray}
&&i\partial_0 \eta_{\alpha}-i\sigma_i\partial_i \eta_{\alpha}
-M_{\alpha\alpha}^* (i\sigma_2\eta_{\alpha}^*)-M_{\beta\alpha}^* (i\sigma_2\eta_{\beta}^*)=0. \label{eom-eta-a}
\end{eqnarray}
In the same way, from the Eular-Lagrange equation for $\overline{\psi_{\alpha L}^c}$, 
\begin{eqnarray}
\frac{\partial L}{\partial \overline{\psi_{\alpha L}^c}}
-\partial_\mu \left(\frac{\partial L}{\partial(\partial_\mu \overline{\psi_{\alpha L}^c})}\right)=0,
\end{eqnarray}
we obtain the equation, 
\begin{eqnarray}
i\gamma^\mu \partial_\mu \psi_{\alpha L}^c-M_{\alpha\alpha} \psi_{\alpha L}-M_{\alpha\beta} \psi_{\beta L}=0.
\end{eqnarray}
Multiplying $\gamma_0$ from the left, the above equation becomes 
\begin{eqnarray}
&&i\partial_0 \psi_{\alpha L}^c+i\gamma^0\gamma^i\partial_i \psi_{\alpha L}^c
-M_{\alpha\alpha} \gamma^0\psi_{\alpha L}-M_{\alpha\beta} \gamma^0\psi_{\beta L}=0.
\end{eqnarray}
Substituting (\ref{psi-def}) and (\ref{nuc}) into this equation, 
we obtain the equation for two-component spinor $\eta_{\alpha}^*$, 
\begin{eqnarray}
i\partial_0 \left(\begin{array}{c}i\sigma_2\eta_{\alpha}^* \\ 0\end{array}\right)+i
\left(\begin{array}{c}\sigma_i\partial_i (i\sigma_2\eta_{\alpha}^*) \\ 0\end{array}\right)
-M_{\alpha\alpha} \left(\begin{array}{c}\eta_{\alpha} \\ 0\end{array}\right)
-M_{\alpha\beta} \left(\begin{array}{c}\eta_{\beta} \\ 0\end{array}\right)=0. 
\end{eqnarray}
Taking out the upper two components, we obtain 
\begin{eqnarray}
&&i\partial_0 (i\sigma_2\eta_{\alpha}^*)+i\sigma_i\partial_i (i\sigma_2\eta_{\alpha}^*)
-M_{\alpha\alpha} \eta_{\alpha}-M_{\alpha\beta} \eta_{\beta}=0. \label{eom-etac-a}
\end{eqnarray}
If we take the equal momentum assumption, 
\begin{eqnarray}
\eta_{\alpha}(x,t)=e^{i\vec{p}\cdot \vec{x}}\eta_{\alpha}(t)
=e^{i\vec{p}\cdot \vec{x}}\left(\begin{array}{c}\nu_{\alpha L}^{\prime} \\ \nu_{\alpha L}\end{array}\right), \quad
\eta_{\beta}(x,t)=e^{i\vec{p}\cdot \vec{x}}\eta_{\beta}(t),
=e^{i\vec{p}\cdot \vec{x}}\left(\begin{array}{c}\nu_{\beta L}^{\prime} \\ \nu_{\beta L}\end{array}\right), 
\end{eqnarray}
and choose $\vec{p}=(0,0,p)$ as momentum vector, 
the complex conjugate of these two-component spinors are given by 
\begin{eqnarray}
\eta_{\alpha}^*(x,t)=e^{-i\vec{p}\cdot \vec{x}}\eta_{\alpha}^*(t)
=e^{-i\vec{p}\cdot \vec{x}}\left(\begin{array}{c}\nu_{\alpha L}^{*\prime} \\ \nu_{\alpha L}^*\end{array}\right), \quad
\eta_{\beta}^*(x,t)=e^{-i\vec{p}\cdot \vec{x}}\eta_{\beta}^*(t)
=e^{-i\vec{p}\cdot \vec{x}}\left(\begin{array}{c}\nu_{\beta L}^{*\prime} \\ \nu_{\beta L}^*\end{array}\right).
\end{eqnarray}
Note that the sign of the momentum for $\eta$ and $\eta^*$ is opposite.
Then, (\ref{eom-eta-a}) and (\ref{eom-etac-a}) are rewritten as 
\begin{eqnarray}
&&\hspace{-1cm}i\partial_0 \left(\begin{array}{c}\nu_{\alpha L}^{\prime} \\ \nu_{\alpha L}\end{array}\right)
+p\left(\begin{array}{c}\nu_{\alpha L}^{\prime} \\ -\nu_{\alpha L}\end{array}\right)
-M_{\alpha\alpha}^* \left(\begin{array}{c}\nu_{\alpha L}^{c} \\ \nu_{\alpha L}^{c\prime}\end{array}\right)
-M_{\beta\alpha}^* \left(\begin{array}{c}\nu_{\beta L}^{c} \\ \nu_{\beta L}^{c\prime}\end{array}\right)=0, \\
&&\hspace{-1cm}i\partial_0 \left(\begin{array}{c}\nu_{\alpha L}^{c} \\ \nu_{\alpha L}^{c\prime}\end{array}\right)
+p\left(\begin{array}{c}\nu_{\alpha L}^{c} \\ -\nu_{\alpha L}^{c\prime}\end{array}\right)
-M_{\alpha\alpha} \left(\begin{array}{c}\nu_{\alpha L}^{\prime} \\ \nu_{\alpha L}\end{array}\right)
-M_{\alpha\beta} \left(\begin{array}{c}\nu_{\beta L}^{\prime} \\ \nu_{\beta L}\end{array}\right)=0, 
\end{eqnarray}
where we replace $\nu^*$ and $\nu^{*\prime}$ to $\nu^c$ and $\nu^{c\prime}$ according to the definition 
(\ref{nuc}).
We put the above two equations together as matrix form, 
\begin{eqnarray}
i\frac{d}{dt}\left(\begin{array}{c}
\nu_{\alpha L}^{c} \\ \nu_{\alpha L}^{c\prime} \\ \nu_{\alpha L}^{\prime} \\ \nu_{\alpha L} \\
\nu_{\beta L}^{c} \\ \nu_{\beta L}^{c\prime} \\ \nu_{\beta L}^{\prime} \\ \nu_{\beta L}
\end{array}\right)
=\left(\begin{array}{cccc|cccc}
-p & 0 & M_{\alpha\alpha} & 0 & 0 & 0 & M_{\alpha\beta} & 0 \\ 
0 & p & 0 & M_{\alpha\alpha} & 0 & 0 & 0 & M_{\alpha\beta} \\
M_{\alpha\alpha}^* & 0 & -p & 0 & M_{\beta\alpha}^* & 0 & 0 & 0 \\ 
0 & M_{\alpha\alpha}^* & 0 & p & 0 & M_{\beta\alpha}^* & 0 & 0 \\
\hline
0 & 0 & M_{\beta\alpha} & 0 & -p & 0 & M_{\beta\beta} & 0 \\
0 & 0 & 0 & M_{\beta\alpha} & 0 & p & 0 & M_{\beta\beta} \\
M_{\alpha\beta}^* & 0 & 0 & 0 & M_{\beta\beta}^* & 0 & -p & 0 \\
0 & M_{\alpha\beta}^* & 0 & 0 & 0 & M_{\beta\beta}^* & 0 & p 
\end{array}\right)\left(\begin{array}{c}
\nu_{\alpha L}^{c} \\ \nu_{\alpha L}^{c\prime} \\ \nu_{\alpha L}^{\prime} \\ \nu_{\alpha L} \\
\nu_{\beta L}^{c} \\ \nu_{\beta L}^{c\prime} \\ \nu_{\beta L}^{\prime} \\ \nu_{\beta L}
\end{array}\right).
\end{eqnarray}
Exchanging some rows and some columns, the above equation can be rewritten as  
\begin{eqnarray}
i\frac{d}{dt}\left(\begin{array}{c}
\nu_{\alpha L}^{c\prime} \\ \nu_{\beta L}^{c\prime} \\ \nu_{\alpha L}^{\prime} \\ \nu_{\beta L}^{\prime} \\
\nu_{\alpha L}^c \\ \nu_{\beta L}^c \\ \nu_{\alpha L} \\ \nu_{\beta L}
\end{array}\right)
=\left(\begin{array}{cccc|cccc}
p & 0 & M_{\alpha\alpha} & M_{\alpha\beta} & 0 & 0 & 0 & 0 \\ 
0 & p & M_{\beta\alpha} & M_{\beta\beta} & 0 & 0 & 0 & 0 \\
M_{\alpha\alpha}^* & M_{\beta\alpha}^* & -p & 0 & 0 & 0 & 0 & 0 \\ 
M_{\alpha\beta}^* & M_{\beta\beta}^* & 0 & -p & 0 & 0 & 0 & 0 \\
\hline
0 & 0 & 0 & 0 & -p & 0 & M_{\alpha\alpha} & M_{\alpha\beta} \\
0 & 0 & 0 & 0 & 0 & -p & M_{\beta\alpha} & M_{\beta\beta} \\
0 & 0 & 0 & 0 & M_{\alpha\alpha}^* & M_{\beta\alpha}^* & p & 0 \\
0 & 0 & 0 & 0 & M_{\alpha\beta}^* & M_{\beta\beta}^* & 0 & p 
\end{array}\right)\left(\begin{array}{c}
\nu_{\alpha L}^{c\prime} \\ \nu_{\beta L}^{c\prime} \\ \nu_{\alpha L}^{\prime} \\ \nu_{\beta L}^{\prime} \\
\nu_{\alpha L}^c \\ \nu_{\beta L}^c \\ \nu_{\alpha L} \\ \nu_{\beta L}
\end{array}\right).
\label{Majorana-Hamiltonian}
\end{eqnarray}
It is noted that $\nu$ and $\nu^c$, which are included in the same multiplet, can transform into each other due to the presence of the Majorana mass term. We can separate the top-left part from the bottom-right part. In the subsequent calculations, we will focus on the bottom-right part, 
\begin{eqnarray}
i\frac{d}{dt}\left(\begin{array}{c}
\nu_{\alpha L}^{c} \\ \nu_{\beta L}^{c} \\
\nu_{\alpha L} \\ \nu_{\beta L}
\end{array}\right)
=\left(\begin{array}{cc|cc}
-p & 0 & M_{\alpha\alpha} & M_{\alpha\beta} \\ 
0 & -p & M_{\beta\alpha} & M_{\beta\beta} \\
\hline
M_{\alpha\alpha}^* & M_{\beta\alpha}^* & p & 0 \\
M_{\alpha\beta}^* & M_{\beta\beta}^* & 0 & p 
\end{array}\right)\left(\begin{array}{c}
\nu_{\alpha L}^{c} \\ \nu_{\beta L}^{c} \\
\nu_{\alpha L} \\ \nu_{\beta L}
\end{array}\right).
\label{Majorana-Hamiltonian2}
\end{eqnarray}
This has the same structure as the equation (\ref{Dirac-Hamiltonian}) discussed in the previous section for Dirac neutrinos. 
The Hamiltonian for Majorana neutrinos is obtained by replacing the Dirac mass with the Majorana mass, corresponding to the substitutions $\nu_{\alpha R} \to \nu_{\alpha L}^{c}$ and $\nu_{\beta R} \to \nu_{\beta L}^{c}$. The Majorana mass matrix becomes a complex symmetric matrix, unlike the Dirac mass matrix, because $\nu^c$ is the complex conjugate of $\nu$ and is not independent of it. The Hamiltonian in eq. (\ref{Majorana-Hamiltonian2}) is then diagonalized as shown in 
\begin{eqnarray}
\hspace{-0.5cm}
\left(\!\!\begin{array}{cc|cc}
U_{\alpha 1} & U_{\beta 1} & 0 & 0 \\
U_{\alpha 2} & U_{\beta 2} & 0 & 0 \\
\hline
0 & 0 & U_{\alpha 1}^* & U_{\beta 1}^* \\
0 & 0 & U_{\alpha 2}^* & U_{\beta 2}^* 
\end{array}\!\!\right)\!\!\!
\left(\!\!\begin{array}{cccc}
-p & 0 & M_{\alpha\alpha} & M_{\alpha\beta} \\
0 & -p & M_{\beta\alpha} & M_{\beta\beta} \\
M_{\alpha\alpha}^* & M_{\beta\alpha}^* & p & 0 \\
M_{\alpha\beta}^* & M_{\beta\beta}^* & 0 & p 
\end{array}\!\!\right)\!\!\!
\left(\!\!\begin{array}{cc|cc}
U_{\alpha 1}^* & U_{\alpha 2}^* & 0 & 0 \\
U_{\beta 1}^* & U_{\beta 2}^* & 0 & 0 \\
\hline
0 & 0 & U_{\alpha 1} & U_{\alpha 2} \\
0 & 0 & U_{\beta 1} & U_{\beta 2} 
\end{array}\!\!\right)\!\!
=\!\!\left(\!\!\begin{array}{cccc}
-p & 0 & m_1 & 0 \\
0 & -p & 0 & m_2 \\
m_1 & 0 & p & 0 \\
0 & m_2 & 0 & p 
\end{array}\!\!\right)\!\!.
\end{eqnarray}
One can see that the $2\times 2$ Majorana mass matrix $M$ of the top-right part in the Hamiltonian 
is diagonalized as $U^TMU={\rm diag}(m_1,m_2)$.
The flavor eigenstates are represented as the linear combination of the 
mass eigenstates, 
\begin{eqnarray}
\left(\begin{array}{c}
\nu_{\alpha L}^{c} \\ \nu_{\beta L}^{c} \\ \nu_{\alpha L} \\ \nu_{\beta L}
\end{array}\right)
=\left(\begin{array}{cc|cc}
U_{\alpha 1}^* & U_{\alpha 2}^* & 0 & 0 \\
U_{\beta 1}^* & U_{\beta 2}^* & 0 & 0 \\
\hline
0 & 0 & U_{\alpha 1} & U_{\alpha 2} \\
0 & 0 & U_{\beta 1} & U_{\beta 2} 
\end{array}\right)\left(\begin{array}{c}
\nu_{1L}^{c} \\ \nu_{2L}^{c} \\ \nu_{1L} \\ \nu_{2L}
\end{array}\right). \label{CFstate-to-CMstate}
\end{eqnarray}
In the case of the Dirac neutrinos, the mass matrix is diagonalized by two unitary matrices $V$ and $U$.
On the other hand, in the case of the Majorana neutrinos, the mass matrix is diagonalized by 
only one unitary matrix $U$.
The time evolution of the mass eigenstates is given by 
\begin{eqnarray}
i\frac{d}{dt}\left(\begin{array}{c}
\nu_{1L}^{c} \\ \nu_{2L}^{c} \\ \nu_{1L} \\ \nu_{2L}
\end{array}\right)
=\left(\begin{array}{cccc}
-p & 0 & m_1 & 0 \\
0 & -p & 0 & m_2 \\
m_1 & 0 & p & 0 \\
0 & m_2 & 0 & p 
\end{array}\right)\left(\begin{array}{c}
\nu_{1L}^{c} \\ \nu_{2L}^{c} \\ \nu_{1L} \\ \nu_{2L}
\end{array}\right).
\end{eqnarray}
Furthermore, let us rewrite this equation to that of the energy eigenstates 
in order to diagonalize the Hamiltonian completely.
Exchanging some rows and some columns in the above equation, we obtain 
\begin{eqnarray}
i\frac{d}{dt}\left(\begin{array}{c}
\nu_{1L}^{c} \\ \nu_{1L} \\ \nu_{2L}^{c} \\ \nu_{2L}
\end{array}\right)
=\left(\begin{array}{cc|cc}
-p & m_1 & 0 & 0 \\
m_1 & p & 0 & 0 \\
\hline
0 & 0 & -p & m_2 \\
0 & 0 & m_2 & p 
\end{array}\right)\left(\begin{array}{c}
\nu_{1L}^{c} \\ \nu_{1L} \\ \nu_{2L}^{c} \\ \nu_{2L}
\end{array}\right). \label{H-for-FMstate}
\end{eqnarray}
The mass eigenstates are represented as the linear combination of 
energy eigenstates, 
\begin{eqnarray}
\left(\begin{array}{c}
\nu_{1L}^{c} \\ \nu_{1L} \\ \nu_{2L}^{c} \\ \nu_{2L}
\end{array}\right)
=\left(\begin{array}{cc|cc}
\sqrt{\frac{E_1+p}{2E_1}} & \sqrt{\frac{E_1-p}{2E_1}} & 0 & 0 \\
-\sqrt{\frac{E_1-p}{2E_1}} & \sqrt{\frac{E_1+p}{2E_1}} & 0 & 0 \\
\hline
0 & 0 & \sqrt{\frac{E_2+p}{2E_2}} & \sqrt{\frac{E_2-p}{2E_2}} \\
0 & 0 & -\sqrt{\frac{E_2-p}{2E_2}} & \sqrt{\frac{E_2+p}{2E_2}} 
\end{array}\right)\left(\begin{array}{c}
\nu_{1}^{c-} \\ \nu_{1}^+ \\ \nu_{2}^{c-} \\ \nu_{2}^+
\end{array}\right), \label{CMstate-to-EHstate}
\end{eqnarray}
by using the matrix diagonalizing the Hamiltonian in eq. (\ref{H-for-FMstate}).
Then, the time evolution of the energy eigenstates is given by 
\begin{eqnarray}
i\frac{d}{dt}\left(\begin{array}{c}
\nu_{1}^{c-} \\ \nu_{1}^+ \\ \nu_{2}^{c-} \\ \nu_{2}^+
\end{array}\right)
=\left(\begin{array}{cccc}
-E_1 & 0 & 0 & 0 \\
0 & E_1 & 0 & 0 \\
0 & 0 & -E_2 & 0 \\
0 & 0 & 0 & E_2 
\end{array}\right)\left(\begin{array}{c}
\nu_{1}^{c-} \\ \nu_{1}^+ \\ \nu_{2}^{c-} \\ \nu_{2}^+
\end{array}\right),
\end{eqnarray}
where 
\begin{eqnarray}
E_1=\sqrt{p^2+m_1^2}, \qquad
E_2=\sqrt{p^2+m_2^2}. 
\end{eqnarray}
Combining the equations, (\ref{CFstate-to-CMstate}) and (\ref{CMstate-to-EHstate}), 
the flavor eigenstates are represented as the linear combination of 
the energy eigenstates, 
\begin{eqnarray}
\left(\begin{array}{c}
\nu_{\alpha L}^{c} \\ \nu_{\beta L}^{c} \\ \nu_{\alpha L} \\ \nu_{\beta L}
\end{array}\right)\!\!
&=&\!\!
\left(\begin{array}{cc|cc}
U_{\alpha 1}^* & U_{\alpha 2}^* & 0 & 0 \\
U_{\beta 1}^* & U_{\beta 2}^* & 0 & 0 \\
\hline
0 & 0 & U_{\alpha 1} & U_{\alpha 2} \\
0 & 0 & U_{\beta 1} & U_{\beta 2} 
\end{array}\right)\!\!\!
\left(\begin{array}{cccc}
1 & 0 & 0 & 0 \\
0 & 0 & 1 & 0 \\
0 & 1 & 0 & 0 \\
0 & 0 & 0 & 1 
\end{array}\right)\!\!\!
\left(\begin{array}{cc|cc}
\sqrt{\frac{E_1+p}{2E_1}} & \sqrt{\frac{E_1-p}{2E_1}} & 0 & 0 \\
-\sqrt{\frac{E_1-p}{2E_1}} & \sqrt{\frac{E_1+p}{2E_1}} & 0 & 0 \\
\hline
0 & 0 & \sqrt{\frac{E_2+p}{2E_2}} & \sqrt{\frac{E_2-p}{2E_2}} \\
0 & 0 & -\sqrt{\frac{E_2-p}{2E_2}} & \sqrt{\frac{E_2+p}{2E_2}} 
\end{array}\right)\!\!\!
\left(\begin{array}{c}
\nu_{1}^{c-} \\ \nu_{1}^+ \\ \nu_{2}^{c-} \\ \nu_{2}^+
\end{array}\right) \nonumber \\
&=&\left(\begin{array}{cc|cc}
\sqrt{\frac{E_1+p}{2E_1}}U_{\alpha 1}^* & \sqrt{\frac{E_1-p}{2E_1}}U_{\alpha 1}^* 
& \sqrt{\frac{E_2+p}{2E_2}}U_{\alpha 2}^* & \sqrt{\frac{E_2-p}{2E_2}}U_{\alpha 2}^* \\
\sqrt{\frac{E_1+p}{2E_1}}U_{\beta 1}^* & \sqrt{\frac{E_1-p}{2E_1}}U_{\beta 1}^* 
& \sqrt{\frac{E_2+p}{2E_2}}U_{\beta 2}^* & \sqrt{\frac{E_2-p}{2E_2}}U_{\beta 2}^* \\
\hline
-\sqrt{\frac{E_1-p}{2E_1}}U_{\alpha 1} & \sqrt{\frac{E_1+p}{2E_1}}U_{\alpha 1} 
& -\sqrt{\frac{E_2-p}{2E_2}}U_{\alpha 2} & \sqrt{\frac{E_2+p}{2E_2}}U_{\alpha 2} \\
-\sqrt{\frac{E_1-p}{2E_1}}U_{\beta 1} & \sqrt{\frac{E_1+p}{2E_1}}U_{\beta 1} 
& -\sqrt{\frac{E_2-p}{2E_2}}U_{\beta 2} & \sqrt{\frac{E_2+p}{2E_2}}U_{\beta 2} 
\end{array}\right)
\left(\begin{array}{c}
\nu_{1}^{c-} \\ \nu_{1}^+ \\ \nu_{2}^{c-} \\ \nu_{2}^+
\end{array}\right). 
\end{eqnarray}
Rewriting the relation for the fields obtained from the above equation into the relation for one particle states, 
we obtain the flavor eigenstates after the time $t$,  
\begin{eqnarray}
&&\hspace{-0.5cm}|\nu_{\alpha L}^{c}(t)\rangle =U_{\alpha 1}\sqrt{\frac{E_1+p}{2E_1}}e^{iE_1t}|\nu_1^{c-}\rangle 
+U_{\alpha 1}\sqrt{\frac{E_1-p}{2E_1}}e^{-iE_1t} |\nu_1^+\rangle \nonumber \\
&&+U_{\alpha 2}\sqrt{\frac{E_2+p}{2E_2}}e^{iE_2t}|\nu_2^{c-}\rangle 
+U_{\alpha 2}\sqrt{\frac{E_2-p}{2E_2}}e^{-iE_2t}|\nu_2^+\rangle, \\
&&\hspace{-0.5cm}|\nu_{\beta L}^{c}(t)\rangle =U_{\beta 1}\sqrt{\frac{E_1+p}{2E_1}}e^{iE_1t}|\nu_1^{c-}\rangle 
+U_{\beta 1}\sqrt{\frac{E_1-p}{2E_1}}e^{-iE_1t}|\nu_1^+\rangle  \nonumber \\
&&+U_{\beta 2}\sqrt{\frac{E_2+p}{2E_2}}e^{iE_2t}|\nu_2^{c-}\rangle 
+U_{\beta 2}\sqrt{\frac{E_2-p}{2E_2}}e^{-iE_2t}|\nu_2^+\rangle,  \\
&&\hspace{-0.5cm}|\nu_{\alpha L}(t)\rangle =-U_{\alpha 1}^*\sqrt{\frac{E_1-p}{2E_1}}e^{iE_1t}|\nu_1^{c-}\rangle 
+U_{\alpha 1}^*\sqrt{\frac{E_1+p}{2E_1}}e^{-iE_1t}|\nu_1^+\rangle  \nonumber \\
&&-U_{\alpha 2}^*\sqrt{\frac{E_2-p}{2E_2}}e^{iE_2t}|\nu_2^{c-}\rangle 
+U_{\alpha 2}^*\sqrt{\frac{E_2+p}{2E_2}}e^{-iE_2t}|\nu_2^+\rangle,  \\
&&\hspace{-0.5cm}|\nu_{\beta L}(t)\rangle =-U_{\beta 1}^*\sqrt{\frac{E_1-p}{2E_1}}e^{iE_1t}|\nu_1^{c-}\rangle 
+U_{\beta 1}^*\sqrt{\frac{E_1+p}{2E_1}}e^{-iE_1t}|\nu_1^+\rangle  \nonumber \\
&&-U_{\beta 2}^*\sqrt{\frac{E_2-p}{2E_2}}e^{iE_2t}|\nu_2^{c-}\rangle 
+U_{\beta 2}^*\sqrt{\frac{E_2+p}{2E_2}}e^{-iE_2t}|\nu_2^+\rangle.  
\end{eqnarray}
and their conjugate states are also given by 
\begin{eqnarray}
&&\hspace{-1.2cm}\langle \nu_{\alpha L}^{c}|\!=\!U_{\alpha 1}^*\sqrt{\frac{E_1+p}{2E_1}}\langle \nu_1^{c-}|\!
+\!U_{\alpha 1}^*\sqrt{\frac{E_1-p}{2E_1}}\langle \nu_1^+|\! 
+\!U_{\alpha 2}^*\sqrt{\frac{E_2+p}{2E_2}}\langle \nu_2^{c-}|\!
+\!U_{\alpha 2}^*\sqrt{\frac{E_2-p}{2E_2}}\langle \nu_2^+|, \\
&&\hspace{-1.2cm}\langle \nu_{\beta L}^{c}|\!=\!U_{\beta 1}^*\sqrt{\frac{E_1+p}{2E_1}}\langle \nu_1^{c-}| \!
+\!U_{\beta 1}^*\sqrt{\frac{E_1-p}{2E_1}}\langle \nu_1^+|\!  
+\!U_{\beta 2}^*\sqrt{\frac{E_2+p}{2E_2}}\langle \nu_2^{c-}|\! 
+\!U_{\beta 2}^*\sqrt{\frac{E_2-p}{2E_2}}\langle \nu_2^+|,  \\
&&\hspace{-1.2cm}\langle \nu_{\alpha L}|\!=\!-U_{\alpha 1}\sqrt{\!\frac{E_1-p}{2E_1}}\langle \nu_1^{c-}|\!
+\!U_{\alpha 1}\sqrt{\!\frac{E_1+p}{2E_1}}\langle \nu_1^+|\!  
-\!U_{\alpha 2}\sqrt{\!\frac{E_2-p}{2E_2}}\langle \nu_2^{c-}|\! 
+\!U_{\alpha 2}\sqrt{\!\frac{E_2+p}{2E_2}}\langle \nu_2^+|,  \\
&&\hspace{-1.2cm}\langle \nu_{\beta L}|\!=\!-U_{\beta 1}\sqrt{\!\frac{E_1-p}{2E_1}}\langle \nu_1^{c-}| \!
+\!U_{\beta 1}\sqrt{\!\frac{E_1+p}{2E_1}}\langle \nu_1^+|\!  
-\!U_{\beta 2}\sqrt{\!\frac{E_2-p}{2E_2}}\langle \nu_2^{c-}|\! 
+\!U_{\beta 2}\sqrt{\!\frac{E_2+p}{2E_2}}\langle \nu_2^+|.  
\end{eqnarray}
From these relations, we obtain the oscillation amplitudes, 
\begin{eqnarray}
&&\hspace{-1cm}A(\nu_{\alpha L}\to\nu_{\alpha L})=\langle \nu_{\alpha L}|\nu_{\alpha L}(t)\rangle \nonumber \\
&&\hspace{-0.7cm}=|U_{\alpha 1}|^2\frac{E_1-p}{2E_1}e^{iE_1t}
+|U_{\alpha 1}|^2\frac{E_1+p}{2E_1}e^{-iE_1t}
+|U_{\alpha 2}|^2\frac{E_2-p}{2E_2}e^{iE_2t}
+|U_{\alpha 2}|^2\frac{E_2+p}{2E_2}e^{-iE_2t} \nonumber \\
&&\hspace{-0.7cm}=|U_{\alpha 1}|^2\cos (E_1t)-i|U_{\alpha 1}|^2\cdot \frac{p}{E_1}\sin (E_1t)
+|U_{\alpha 2}|^2\cos (E_2t)-i|U_{\alpha 2}|^2\cdot \frac{p}{E_2}\sin (E_2t), \\
&&\hspace{-1cm}A(\nu_{\alpha L}\to\nu_{\beta L})=\langle \nu_{\beta L}|\nu_{\alpha L}(t)\rangle \nonumber \\
&&\hspace{-0.7cm}=U_{\alpha 1}^*U_{\beta 1}\left(\frac{E_1-p}{2E_1}e^{iE_1t}+\frac{E_1+p}{2E_1}e^{-iE_1t}\right)
+U_{\alpha 2}^*U_{\beta 2}\left(\frac{E_2-p}{2E_2}e^{iE_2t}+\frac{E_2+p}{2E_2}e^{-iE_2t}\right) \nonumber \\
&&\hspace{-0.7cm}=U_{e1}^*U_{\mu 1}\left(\cos (E_1t)-i\frac{p}{E_1}\sin (E_1t)\right)
+U_{\alpha 2}^*U_{\beta 2}\left(\cos (E_2t)-i\frac{p}{E_2}\sin (E_2t)\right), \\
&&\hspace{-1cm}A(\nu_{\alpha L}\to\nu_{\alpha L}^{c})=\langle \nu_{\alpha L}^c|\nu_{\alpha L}(t)\rangle \nonumber \\
&&\hspace{-0.7cm}=-U_{\alpha 1}^*U_{\alpha 1}^*\frac{m_1}{2E_1}(e^{iE_1t}-e^{-iE_1t})
-U_{\alpha 2}^*U_{\alpha 2}^*\frac{m_2}{2E_2}(e^{iE_2t}-e^{-iE_2t}) \nonumber \\
&&\hspace{-0.7cm}=-iU_{\alpha 1}^{*2}\frac{m_1}{E_1}\sin (E_1t)-iU_{\alpha 2}^{*2}\frac{m_2}{E_2}\sin (E_2t), \\
&&\hspace{-1cm}A(\nu_{\alpha L}\to\nu_{\beta L}^{c})=\langle \nu_{\beta L}^c|\nu_{\alpha L}(t)\rangle \nonumber \\
&&\hspace{-0.7cm}=U_{\alpha 1}^*U_{\beta 1}^*\frac{m_1}{E_1}(e^{iE_1t}-e^{-iE_1t})
-U_{\alpha 2}^*U_{\beta 2}^*\frac{m_2}{E_2}(e^{iE_2t}-e^{-iE_2t})\nonumber \\
&&\hspace{-0.7cm}=-iU_{\alpha 1}^*U_{\beta 1}^*\frac{m_1}{E_1}\sin (E_1t)
-iU_{\alpha 2}^*U_{\beta 2}^*\frac{m_2}{E_2}\sin (E_2t). 
\end{eqnarray}
Furthermore, we calculate the oscillation probabilities for $\nu_{\alpha L}$ by squaring the corresponding amplitudes, 
\begin{eqnarray}
&&\hspace{-1cm}P(\nu_{\alpha L}\to\nu_{\alpha L})=
[|U_{\alpha 1}|^2\cos (E_1t)+|U_{\alpha 2}|^2\cos (E_2t)]^2 \nonumber \\
&&\hspace{-0.3cm}+\left[|U_{\alpha 1}|^2\cdot \frac{p}{E_1}\sin (E_1t)
+|U_{\alpha 2}|^2\cdot \frac{p}{E_2}\sin (E_2t)\right]^2, \label{P1} \\
&&\hspace{-1cm}P(\nu_{\alpha L}\to\nu_{\beta L})=
|U_{\alpha 1}U_{\beta 1}|^2\left\{\cos^2 (E_1t)+\frac{p^2}{E_1^2}\sin^2 (E_1t)\right\}  \nonumber \\
&&\hspace{-0.3cm}+|U_{\alpha 2}U_{\beta 2}|^2\left\{\cos^2 (E_2t)+\frac{p^2}{E_2^2}\sin^2 (E_2t)\right\}\nonumber \\
&&\hspace{-0.3cm}+2{\rm Re}\left[U_{\alpha 1}^*U_{\beta 1}\left(\cos (E_1t)-i\frac{p}{E_1}\sin (E_1t)\right)
U_{\alpha 2}U_{\beta 2}^*\left(\cos (E_2t)+i\frac{p}{E_2}\sin (E_2t)\right)\right], \\
&&\hspace{-1cm}P(\nu_{\alpha L}\to\nu_{\alpha L}^{c})=
|U_{\alpha 1}^*U_{\alpha 1}^*|^2\frac{m_1^2}{E_1^2}\sin^2 (E_1t)+|U_{\alpha 2}^*U_{\alpha 2}^*|^2\frac{m_2^2}{E_2^2}
\sin^2 (E_2t)  \nonumber \\
&&\hspace{-0.3cm}+2{\rm Re}(U_{\alpha 1}^*U_{\alpha 2}U_{\alpha 1}^*U_{\alpha 2})\frac{m_1m_2}{E_1E_2}\sin (E_1t)\sin (E_2t), \\
&&\hspace{-1cm}P(\nu_{\alpha L}\to\nu_{\beta L}^{c})=
|U_{\alpha 1}^*U_{\beta 1}^*|^2\frac{m_1^2}{E_1^2}\sin^2 (E_1t)
+|U_{\alpha 2}^*U_{\beta 2}^*|^2\frac{m_2^2}{E_2^2}\sin^2 (E_2t)  \nonumber \\
&&\hspace{-0.3cm}+2{\rm Re}(U_{\alpha 1}^*U_{\alpha 2}U_{\beta 1}^*U_{\beta 2})\frac{m_1m_2}{E_1E_2}\sin (E_1t)\sin (E_2t). \label{P4}
\end{eqnarray}
As $2\times 2$ unitary matrix has four parameters in general, we choose the following 
parametrization,  
\begin{eqnarray}
U&=&\left(\begin{array}{cc}
e^{i\rho_1} & 0 \\ 
0 & e^{i\rho_2} 
\end{array}\right)
\left(\begin{array}{cc}
\cos \theta & \sin \theta \\ 
-\sin \theta & \cos \theta 
\end{array}\right)\left(\begin{array}{cc}
1 & 0 \\ 
0 & e^{i\phi} 
\end{array}\right)
=
\left(\begin{array}{cc}
e^{i\rho_1}\cos \theta & e^{i(\rho_1+\phi)}\sin \theta \\ 
-e^{i\rho_2}\sin \theta & e^{i(\rho_2+\phi)}\cos \theta 
\end{array}\right).
\end{eqnarray}
Substituting these parametrization into the equations (\ref{P1})-(\ref{P4}), 
the oscillation probabilities are rewritten as 
\begin{eqnarray}
&&\hspace{-1cm}P(\nu_{\alpha L}\to\nu_{\alpha L})=
1-4s^2c^2\sin^2 \frac{(E_2-E_1)t}{2} \nonumber \\
&&\hspace{-0.5cm}-\left[c^4\cdot \frac{m_1^2}{E_1^2}\sin^2 (E_1t)
+s^4\cdot \frac{m_2^2}{E_2^2}\sin^2 (E_2t)+2s^2c^2 \left(1-\frac{p^2}{E_1E_2}\right)
\sin (E_1t)\sin (E_2t)\right], \\
&&\hspace{-1cm}P(\nu_{\alpha L}\to\nu_{\beta L})=
4s^2c^2\sin^2 \frac{(E_2-E_1)t}{2}\nonumber \\
&&\hspace{-0.5cm}-s^2c^2\left[\frac{m_1^2}{E_1^2}\sin^2 (E_1t)+\frac{m_2^2}{E_2^2}\sin^2 (E_2t)
-2\left(1-\frac{p^2}{E_1E_2}\right)\sin (E_1t)\sin (E_2t)\right],\\
&&\hspace{-1cm}P(\nu_{\alpha L}\to\nu_{\alpha L}^{c})=
c^4\frac{m_1^2}{E_1^2}\sin^2 (E_1t)+s^4\frac{m_2^2}{E_2^2}\sin^2 (E_2t) \nonumber \\
&&\hspace{-0.5cm}+2s^2c^2\cos(2\phi)\frac{m_1m_2}{E_1E_2}\sin (E_1t)\sin (E_2t), \label{P3b} \\
&&\hspace{-1cm}P(\nu_{\alpha L}\to\nu_{\beta L}^{c})=
c^2s^2\frac{m_1^2}{E_1^2}\sin^2 (E_1t)
+s^2c^2\frac{m_2^2}{E_2^2}\sin^2 (E_2t) \nonumber \\
&&\hspace{-0.5cm}-2s^2c^2\cos(2\phi)\frac{m_1m_2}{E_1E_2}\sin (E_1t)\sin (E_2t). \label{P4b}
\end{eqnarray}
The probabilities for the Majorana neutrinos are obtained by the replacements, 
$\theta_L=\theta_R=\theta$, $\phi_L=-\phi_R=\phi$ in the probabilities for the Dirac neutrinos.
This reflects that $\nu_L^c$ plays the role of the right-handed components of the Dirac neutrinos $\nu_R$,  
in the case of the Majorana neutrinos.
It has been considered that the Majorana CP phase appears accompanied by the lepton number violation until now.
However, we have shown that a new CP phase appears even in the case of two-generation Dirac neutrinos and non-existence of lepton number violation in our previous papers. 
(This new CP phase is absorbed by the redefinition of the fields in the Standard Model because the right-handed 
neutrinos have no weak interactions.) 
In the case of the Majorana neutrinos, $\nu_L^c$, an alternative of the right-handed neutrino, 
has weak interactions. As the result, this new CP phase appears in the oscillation probabilities as the Majorana CP phase.
Therefore, we can reinterpret that the Majorana CP phase is accompanied by chirality change rather than lepton number violation. 
Also note that the sign of the momentum changes in $\nu \leftrightarrow \nu^c$ oscillations, 
namely the probabilities in (\ref{P3b}) and (\ref{P4b}) mean $P(\nu_{\alpha L}(p) \to \nu_{\alpha L}^c (-p))$ 
and $P(\nu_{\alpha L}(p) \to \nu_{\beta L}^c (-p))$, respectively. 

Next, let us derive the oscillation probabilities of antineutrino $\nu_{\alpha L}^c(p)$ with momentum $p$.
In eq.(\ref{Majorana-Hamiltonian2}), $\nu_L^c$ has negative momentum $-p$. In order to derive 
the oscillation probabilities for $\nu_{\alpha L}^c(p)$, 
we need to change the sign of momentum $p$ in eq.(\ref{Majorana-Hamiltonian2}).
Namely, we start from the equation, 
\begin{eqnarray}
i\frac{d}{dt}\left(\begin{array}{c}
\nu_{\alpha L}^{c} \\ \nu_{\beta L}^{c} \\
\nu_{\alpha L} \\ \nu_{\beta L}
\end{array}\right)
=\left(\begin{array}{cc|cc}
p & 0 & M_{\alpha\alpha} & M_{\alpha\beta} \\ 
0 & p & M_{\beta\alpha} & M_{\beta\beta} \\
\hline
M_{\alpha\alpha}^* & M_{\beta\alpha}^* & -p & 0 \\
M_{\alpha\beta}^* & M_{\beta\beta}^* & 0 & -p 
\end{array}\right)\left(\begin{array}{c}
\nu_{\alpha L}^{c} \\ \nu_{\beta L}^{c} \\
\nu_{\alpha L} \\ \nu_{\beta L}
\end{array}\right).
\end{eqnarray}
Furthermore, exchanging the first two rows and the last two rows, we obtain 
\begin{eqnarray}
i\frac{d}{dt}\left(\begin{array}{c}
\nu_{\alpha L} \\ \nu_{\beta L} \\
\nu_{\alpha L}^c \\ \nu_{\beta L}^{c}
\end{array}\right)
=\left(\begin{array}{cc|cc}
-p & 0 & M_{\alpha\alpha}^* & M_{\alpha\beta}^* \\ 
0 & -p & M_{\beta\alpha}^* & M_{\beta\beta}^* \\
\hline
M_{\alpha\alpha} & M_{\beta\alpha} & p & 0 \\
M_{\alpha\beta} & M_{\beta\beta} & 0 & p 
\end{array}\right)\left(\begin{array}{c}
\nu_{\alpha L} \\ \nu_{\beta L} \\
\nu_{\alpha L}^{c} \\ \nu_{\beta L}^{c}
\end{array}\right),
\end{eqnarray}
where we used $M_{\alpha\beta}=M_{\beta\alpha}$ due to the symmetry of the mass matrix.
Comparing with (\ref{Majorana-Hamiltonian2}), one can see the mass terms in the Hamiltonian have changed to 
their complex conjugate.
Therefore, we can obtain the oscillation probabilities of antineutrinos by replacing the mixing matrix 
that diagonalizes the mass matrix with its complex conjugate. 
In two generations, the CP conjugate probabilities are the same as the original probabilities 
because it depends on the absolute value or real part of the product of $U$s.
The probabilities for oscillations with chirality-flip depend on the new CP phase only through the form of $\cos 2\phi$.  
Thus, we obtain the following relations, 
\begin{eqnarray}
P(\nu_{\alpha L}^{c}(p)\to\nu_{\alpha L}^{c}(p))&=&P(\nu_{\alpha L}(p)\to\nu_{\alpha L}(p)), \\
P(\nu_{\alpha L}^{c}(p)\to\nu_{\beta L}^{c}(p))&=&P(\nu_{\alpha L}(p)\to\nu_{\beta L}(p)), \\
P(\nu_{\alpha L}^{c}(p)\to\nu_{\alpha L}(-p))&=&P(\nu_{\alpha L}(p)\to\nu_{\alpha L}^{c}(-p)), \\
P(\nu_{\alpha L}^{c}(p)\to\nu_{\beta L}(-p))&=&P(\nu_{\alpha L}(p)\to\nu_{\beta L}^{c}(-p)).
\end{eqnarray}

\section{Difference between Conventional and New Probabilities}

In this section, we review the results concerning Majorana neutrinos presented in previous works \cite{Li1982, Gouvea2003, Xing2013} and compare these findings with those obtained in the current paper. 
In previous studies, the amplitude for the transition from $\nu_{\alpha L}$ to $\nu_{\beta L}^{c}$, as well as the CP-conjugate amplitude for the transition from $\nu_{\alpha L}^{c}$ to $\nu_{\beta L}$, were given by 
\begin{eqnarray}
A(\nu_{\alpha L}\to \nu_{\beta L}^{c})&=&\left[U_{\alpha 1}^*U_{\beta 1}^*\frac{m_1}{E_1}e^{-iE_1t}+
U_{\alpha 2}^*U_{\beta 2}^*\frac{m_2}{E_2}e^{-iE_2t}\right]K, \\
A(\nu_{\alpha L}^{c}\to \nu_{\beta L})&=&\left[U_{\alpha 1}U_{\beta 1}\frac{m_1}{E_1}e^{-iE_1t}
+U_{\alpha 2}U_{\beta 2}\frac{m_2}{E_2}e^{-iE_2t}\right]\bar{K}, 
\end{eqnarray}
in our notation, where $K=\bar{K}$ is the kinetic factor and does not depend on energy and time 
(namely corresponding distance).
On the other hand, the amplitudes obtained in this paper are represented as 
\begin{eqnarray}
A(\nu_{\alpha L}\to\nu_{\beta L}^{c})&=&U_{\alpha 1}^*U_{\beta 1}^*\frac{m_1}{2E_1}(e^{-iE_1t}-e^{iE_1t})
+U_{\alpha 2}^*U_{\beta 2}^*\frac{m_2}{2E_2}(e^{-iE_2t}-e^{iE_2t})\nonumber \\
&=&-iU_{\alpha 1}^*U_{\beta 1}^*\frac{m_1}{E_1}\sin (E_1t)
-iU_{\alpha 2}^*U_{\beta 2}^*\frac{m_2}{E_2}\sin (E_2t), \\
A(\nu_{\alpha L}^{c}\to\nu_{\beta L})&=&U_{\alpha 1}U_{\beta 1}\frac{m_1}{2E_1}(e^{-iE_1t}-e^{iE_1t})
+U_{\alpha 2}U_{\beta 2}\frac{m_2}{2E_2}(e^{-iE_2t}-e^{iE_2t}) \nonumber \\
&=&-iU_{\alpha 1}U_{\beta 1}\frac{m_1}{E_1}\sin (E_1t)-iU_{\alpha 2}U_{\beta 2}\frac{m_2}{E_2}\sin (E_2t). 
\end{eqnarray}
The difference between conventional result and our result is in the exponential part.
Our result has a contribution coming from both positive and negative energies.
When we use the Dirac equation and derive the oscillation probabilities for $\nu$ and $\nu^c$ in a unified way, 
we need to take into account the contributions from both signs of energy.

In previous papers, it has been considered that the unitarity holds in the probabilities for only 
$\nu \leftrightarrow \nu$ oscillations.
If this is correct, the unitarity does not hold when $\nu \leftrightarrow \nu^c$ oscillations are taken into account 
in addition to the probabilities between neutrinos.
In order to resolve this paradox, we consider both neutrino and antineutrino 
in a unified way. Actually, the total sum of the probabilities for a neutrino is kept at one by adding some correction terms, namely, unitarity holds.

Next, we compare the oscillation probabilities. The probabilities in previous papers were given by 
\begin{eqnarray}
&&\hspace{-0.7cm}P(\nu_{\alpha L}\to \nu_{\beta L}^{c})=
|K|^2\left|U_{\alpha 1}^*U_{\beta 1}^*\frac{m_1}{E_1}e^{-iE_1t}+
U_{\alpha 2}^*U_{\beta 2}^*\frac{m_2}{E_2}e^{-iE_2t}\right|^2 \nonumber \\
&&\hspace{-0.7cm}=|K|^2\left[\left|\frac{m_1^2}{E_1^2} U_{\alpha 1}^*U_{\beta 1}^*\right|^2
+\left|\frac{m_2^2}{E_2^2} U_{\alpha 2}^*U_{\beta 2}^*\right|^2
+\frac{2m_1m_2}{E_1E_2}{\rm Re}\left(U_{\alpha 1}^*U_{\beta 1}^*U_{\alpha 2}U_{\beta 2}e^{-i(E_1-E_2)t}\right)\right] \nonumber \\
&&\hspace{-0.7cm}=|K|^2\left[\left|\frac{m_1^2}{E_1^2} U_{\alpha 1}U_{\beta 1}\right|^2
+\left|\frac{m_2^2}{E_2^2} U_{\alpha 2}U_{\beta 2}\right|^2\right. \nonumber \label{P-in-pw1} \\
&&\hspace{-0.7cm}\left.+\frac{2m_1m_2}{E_1E_2}\!\!\left\{\!{\rm Re}\left(U_{\alpha 1}U_{\beta 1}U_{\alpha 2}^*U_{\beta 2}^*\right)
\cos(E_1-E_2)t 
-{\rm Im}\left(U_{\alpha 1}U_{\beta 1}U_{\alpha 2}^*U_{\beta 2}^*\right)\sin(E_1-E_2)t\right\}\!\!\right], \label{P-in-pw2}
\end{eqnarray}
where we parametrize the MNS matrix, 
\begin{eqnarray}
U&=&
\left(\begin{array}{cc}
e^{i\rho_1}\cos \theta & e^{i(\rho_1+\phi)}\sin \theta \\ 
-e^{i\rho_2}\sin \theta & e^{i(\rho_2+\phi)}\cos \theta 
\end{array}\right), 
\end{eqnarray}
and substituting into (\ref{P-in-pw2}), the probability is rewritten as 
\begin{eqnarray}
&&\hspace{-0.3cm}P(\nu_{\alpha L}\to \nu_{\beta L}^{c}) \nonumber \\
&&\hspace{-0.3cm}=|K|^2c^2s^2\left[\frac{m_1^2}{E_1^2}+\frac{m_2^2}{E_2^2}
-\frac{2m_1m_2}{E_1E_2}\left\{{\rm Re}\left(e^{-2i\phi}\right)
\cos(E_1-E_2)t 
-{\rm Im}\left(e^{-2i\phi}\right)\sin(E_1-E_2)t\right\}\right] \nonumber \\
&&\hspace{-0.3cm}=|K|^2c^2s^2\left[\frac{m_1^2}{E_1^2}+\frac{m_2^2}{E_2^2}
-\frac{2m_1m_2}{E_1E_2}\left\{\cos (2\phi)\cos(E_1-E_2)t 
+\sin (2\phi)\sin(E_1-E_2)t\right\}\right]. \label{P-in-pw3}
\end{eqnarray}
In the same way, we also obtain 
\begin{eqnarray}
&&\hspace{-0.3cm}P(\nu_{\alpha L}^{c}\to \nu_{\beta L}) \nonumber \\
&&\hspace{-0.3cm}=|K|^2c^2s^2\left[\frac{m_1^2}{E_1^2}+\frac{m_2^2}{E_2^2}
-\frac{2m_1m_2}{E_1E_2}\left\{{\rm Re}\left(e^{-2i\phi}\right)
\cos(E_1-E_2)t 
+{\rm Im}\left(e^{-2i\phi}\right)\sin(E_1-E_2)t\right\}\right] \nonumber \\
&&\hspace{-0.3cm}=
|K|^2c^2s^2\left[\frac{m_1^2}{E_1^2}+\frac{m_2^2}{E_2^2}
-\frac{2m_1m_2}{E_1E_2}\left\{\cos (2\phi)\cos(E_1-E_2)t 
-\sin (2\phi)\sin(E_1-E_2)t\right\}\right]. \label{P-in-pw4}
\end{eqnarray}
From the probabilities (\ref{P-in-pw3}) and (\ref{P-in-pw4}), the difference in previous papers 
is given by 
\begin{eqnarray}
P(\nu_{\alpha L}\to\nu_{\beta L}^{c})-P(\nu_{\alpha L}^{c}\to\nu_{\beta L})
=\frac{4|K|^2m_1m_2c^2s^2}{E_1E_2}\sin (2\phi)\sin(E_1-E_2)t.
\end{eqnarray}

On the other hand, we obtain a different result from the previous works, in this paper. 
We have the same probabilities 
\begin{eqnarray}
&&P(\nu_{\alpha L}\to\nu_{\beta L}^{c})=P(\nu_{\alpha L}^{c}\to\nu_{\beta L})\nonumber \\
&&=c^2s^2\left[\frac{m_1^2}{E_1^2}\sin^2 (E_1t)
+\frac{m_2^2}{E_2^2}\sin^2 (E_2t)
-\frac{2m_1m_2}{E_1E_2}\cos(2\phi)\sin (E_1t)\sin (E_2t)\right], \label{P-in-ow}
\end{eqnarray}
for both $\nu_{\alpha L}\to\nu_{\beta L}^{c}$ and $\nu_{\alpha L}^{c}\to\nu_{\beta L}$ 
oscillations.
So, there is no difference between CP-conjugated probabilities, and  
\begin{eqnarray}
P(\nu_{\alpha L}\to\nu_{\beta L}^{c})-P(\nu_{\alpha L}^{c}\to\nu_{\beta L})=0
\end{eqnarray}
is obtained in vacuum. 

Next, let us consider the the limit $L \to 0$ ($t \to 0$).
Taking the limit $t\to 0$ in (\ref{P-in-pw3}), the probability for $\nu_{\alpha L}\to \nu_{\beta L}^{c}$ 
oscillations converges to 
\begin{eqnarray}
P(\nu_{\alpha L}\to \nu_{\beta L}^{c})=|K|^2c^2s^2\left[\frac{m_1^2}{E_1^2}+\frac{m_2^2}{E_2^2}
-\frac{2m_1m_2}{E_1E_2}\cos (2\phi)\right],
\end{eqnarray}
and has a finite value, namely the zero-distance effect appears. 
This effect has been noted because it does not exist in $\nu \leftrightarrow \nu$ oscillations, 
and unique to $\nu \leftrightarrow \nu^c$ oscillations.
However, we have shown that there appears no zero-distance effect even in $\nu \leftrightarrow \nu^c$  
oscillations as same as $\nu \leftrightarrow \nu$ oscillations by taking the limit $t\to 0$ in eq.(\ref{P-in-ow}).

\section{Summary}

We have derived the oscillation probabilities for neutrinos and antineutrinos in a unified manner using a relativistic approach for the case of two-generation Majorana neutrinos. In this context, $\nu_L^c$ plays the role of $\nu_R$ in Dirac neutrinos, and the new CP phase that appears in Dirac neutrino oscillations \cite{KT1} becomes the Majorana CP phase. Consequently, we have discovered that the Majorana CP phase emerges not from lepton number violation but from chirality-flip processes. Moreover, we have confirmed that unitarity holds when considering both neutrinos and antineutrinos.

In our calculations, we find that there is no direct CP violation associated with the sine term of the Majorana CP phase, even in oscillations involving different flavors of neutrinos and antineutrinos. The difference between CP-conjugated probabilities vanishes, and the probabilities depend solely on the cosine term of the Majorana CP phase. Furthermore, we have demonstrated that there is no zero-distance effect, whereby a neutrino instantaneously converts into an antineutrino. These results differ from those reported in previous studies \cite{Li1982, Gouvea2003, Xing2013}.

\section*{Note added in proof} 

We learned that the literature \cite{Li2024} was published in 2024 after we submitted our paper to arXiv in 2021. The authors of this paper derived the oscillation probabilities for Majorana neutrinos using the same method as ours. The vacuum oscillation probabilities, Eq.(19), they derived are the same as the results Eqs.(84)-(87) in this paper, both in the presence and absence of chirality-flips. Furthermore, they extended this method to cases with matter effects, pointing out that new resonance effects emerge related to helicity states, which are distinct from the usual MSW effect. They also noted that in cases where non-relativistic effects become significant, such as with cosmic background neutrinos, this matter effect may provide a possibility for measuring the Majorana CP phase.

\end{document}